\documentclass[onecolumn,preprintnumbers,eqsecnum,superscriptaddress,12pt]{revtex4}
\usepackage{amsfonts}
\usepackage{amssymb}
\usepackage{amsmath}
\usepackage{epsfig}
\usepackage{graphicx}
\usepackage{dcolumn}
\usepackage{bm}

\usepackage{color}

\def\nn{{\nonumber}}

\newcommand{\beq}{\begin{equation}}
\newcommand{\eeq}{\end{equation}}
\newcommand{\beqs}{\begin{eqnarray}}
\newcommand{\eeqs}{\end{eqnarray}}

\newcommand{\be}{\begin{equation}}
\newcommand{\ee}{\end{equation}}
\newcommand{\bea}{\begin{eqnarray}}
\newcommand{\eea}{\end{eqnarray}}

\begin{document}

\title{Chern-Simons effect on the dual hydrodynamics in the Maxwell-Gauss-Bonnet gravity}

\author{Ya-Peng Hu}\email{huyp@pku.edu.cn}
\address{Center for High-Energy Physics, Peking University, Beijing 100871, China}
\address{Center for Quantum Spacetime, Sogang University, Seoul 121-742, Korea}

\author{Chanyong Park}\email{cyong21@sogang.ac.kr}
\address{Center for Quantum Spacetime, Sogang University, Seoul 121-742, Korea}

\begin{abstract}
Following the previous work 1103.3773, we give a more general and systematic discussion on the Chern-Simons effect
in the 5-dimensional Maxwell-Gauss-Bonnet gravity.
 After constructing the first order perturbative black brane solution, we extract the stress tensor and charge current of dual fluid.
From these results, we find out the dependence of some transport coefficients on the Gauss-Bonnet coupling $\alpha$ and Chern-Simons coupling $\kappa_{cs}$.
We also show that the new anomalous term can provide an additional contribution to the anomalous chiral magnetic conductivity.
\end{abstract}

\maketitle

\vspace*{1.cm}

\newpage

\section{Introduction}
According to the AdS/CFT correspondence~\cite{Maldacena:1997re,Gubser:1998bc,Witten:1998qj,Aharony:1999ti}, the
gravitational theory with an asymptotic AdS spacetime can be
reinterpreted in terms of a quantum field theory defined on its boundary.
Particularly, the fluctuations of a classical gravitational theory can be mapped into
some operators in the strongly interacting dual quantum field theory.
In the last decades, the AdS/CFT
correspondence has been widely used to investigate the strongly interacting quantum field theory, and provided important new insights to several fields such as QCD, superconductor, hydrodynamics~\cite{Herzog:2009xv,CasalderreySolana:2011us,Policastro:2001yc,Kovtun:2003wp,Buchel:2003tz,Kovtun:2004de}.

In this paper, we focus on the hydrodynamics where the AdS/CFT correspondence is used to describe
the hydrodynamic behavior of the dual quantum field theory.
The hydrodynamics
can be viewed as an effective description of an interacting quantum
field theory in the long wave-length limit, i.e. the length
scales under consideration are much larger than the correlation
length of the quantum field theory~\cite{Policastro:2001yc,Kovtun:2003wp,Buchel:2003tz,Kovtun:2004de}. Note that, recently a more systematic study of the hydrodynamics via AdS/CFT correspondence named as the Fluid/Gravity correspondence has been proposed~\cite{Bhattacharyya:2008jc}. Through this systematic way, the stress-energy tensor
of the fluid was constructed order by order with the derivative expansion of the
gravity solution. Furthermore, the shear viscosity $\eta$, entropy density
$s$, and their ratio
$\eta/s$ have been also calculated from the first order stress-energy
tensor~\cite{Hur:2008tq,Erdmenger:2008rm,Banerjee:2008th,Tan:2009yg,
Torabian:2009qk,Hu:2010sn}, which exactly agree with the previous hydrodynamic results
obtained from the Kubo formula~\cite{Policastro:2001yc,Kovtun:2003wp,Buchel:2003tz,Kovtun:2004de}.
Besides the stress-energy tensor, the
Fluid/Gravity correspondence was used to investigate the charge current of the boundary fluid by
adding a Maxwell field
to the gravity theory, in which the information of the
thermal conductivity and electrical conductivity were extracted
\cite{Hur:2008tq,Erdmenger:2008rm,Banerjee:2008th,Tan:2009yg,Son:2009tf,Hu:2011ze,Kalaydzhyan:2011vx}.
It is worthwhile noticing that new effect such as anomalous vortical effect can be brought into the hydrodynamics after adding a Chern-Simons term~\cite{Erdmenger:2008rm,Banerjee:2008th,Son:2009tf,Hu:2011ze,Kalaydzhyan:2011vx}. The effect of Chern-Simons term was first discussed in three dimension where the Chern-Simons term makes the Maxwell theory massive ~\cite{Deser:1982vy}. Furthermore, it was also shown that the Chern-Simons term can affect the holographic superconductors in 4 dimension~\cite{Tallarita:2010vu} and the instability of black hole in the five-dimensional Maxwell theory~\cite{Nakamura:2009tf}.  One of our motivations in this paper is to study more systematically the hydrodynamics  including the Chern-Simons
effect via the Fluid/Gravity correspondence.

In order to consider the  more general case, we take a Maxwell-Gauss-Bonnet (MGB) gravity because the Maxwell-Einstein gravity corresponds to a special case of MGB gravity.
In this background,
it was shown that the shear viscosity is different from one calculated in the Maxwell-Einstein gravity~\cite{Brigante:2007nu,Brigante:2008gz,KP,Brustein:2008cg,Buchel:2009sk,deBoer:2009pn,Camanho:2009vw,Cremonini:2009sy,Ge:2008ni}. In this paper, we show that
other transport coefficients like the
charge diffusion constant and electric conductivity also depends nontrivially on the GB coupling. As
the GB coupling increases, those transport coefficients are also increases.
In addition, we also find that the nonzero external gauge field $A_{\mu}^{\text ext}$
can provide an additional contribution to the chiral magnetic conductivity through the
anomalous Chern-Simons term. In the previous works \cite{Erdmenger:2008rm,Banerjee:2008th,Son:2009tf,Hu:2011ze,Kalaydzhyan:2011vx}, the external gauge field was set
to zero or slowly deviated from zero.

The rest of our paper is organized as follows. In Sec.~II, we
perform a more general and systematic study for effects of the
Chern-Simons and Gauss-Bonnet term on the hydrodynamics.
In the Maxwell-Gauss-Bonnet gravity with the Chern-Simons term,
we calculate the several first order
transport coefficients depending on the Gauss-Bonnet or Chern-Simons coupling .
In Sec.~III, we finish this work with some concluding remarks.

\section{The Chern-Simons effect on the hydrodynamics via AdS/CFT correspondence}

In Refs~\cite{Erdmenger:2008rm,Banerjee:2008th,Son:2009tf,Hu:2011ze,Kalaydzhyan:2011vx}, there were some discussions about the Chern-Simons effects on the hydrodynamics. In this section, we will provide a further systematic discussion on the Chern-Simons effects following our previous work~\cite{Hu:2011ze}.

The action of the 5-dimensional MGB gravity with Chern-Simons term is given by
\begin{equation}
\label{action1} I=\frac{1}{16 \pi G}\int_\mathcal{M}~d^5x \sqrt{-g^{(5)}}
\left(R-2 \Lambda+\alpha L_{GB} \right)-\frac{1}{4g^2}\int_\mathcal{M}~d^5x \sqrt{-g^{(5)}}(F^2+\frac{4\kappa_{cs} }{3}\epsilon ^{LABCD}A_{L}F_{AB}F_{CD}),
\end{equation}
and the equations of motion become
\begin{eqnarray}
\label{eqs1}
R_{AB } -\frac{1}{2}Rg_{AB}+\Lambda g_{AB}+\alpha H_{AB}-\frac{1}{2g^2}\left(F_{A C}{F_{B }}^{C}-\frac{1}{4}g_{AB}F^2\right)&=&0~, \\
\nabla_{B} {F^{B}}_{A}-\kappa_{cs} \epsilon_{ABCDE}F^{BC}F^{DE} &=&0,\nonumber~~
\end{eqnarray}
where we have set $16 \pi G =1$ for simplicity
and $\alpha$ represents a GB coupling with the  $(length)^2$ dimension.
The GB term $L_{GB}$ is
\begin{eqnarray}
\label{LGB} L_{GB} =R^2-4R_{AB}R^{AB}+R_{ABCD}R^{ABCD},
\end{eqnarray}
and $H_{AB}$ implies
\begin{equation}
\label{Hmn}
H_{AB}=2(R_{ALCD }R_{B }^{\phantom{\nu}%
LCD }-2R_{A C B D }R^{CD
}-2R_{A
C }R_{\phantom{\sigma}B }^{C }+RR_{AB })-\frac{1}{2}%
L_{GB}g_{AB }  ~.
\end{equation}
From (\ref{eqs1}), the boosted black brane solution is
given by ~\cite{Cvetic:2001bk,Anninos:2008sj}
\begin{eqnarray}   \label{rnboost}
ds^2 &=& - r^2 f(r)( u_\mu dx^\mu )^2 - 2 u_\mu dx^\mu dr + \frac{r^2}{\ell_c^2}P_{\mu \nu} dx^\mu dx^\nu, \\
f(r) &=& \frac{1}{4\alpha}
 \bigg (
 1-\sqrt{1-8\alpha (1-\frac{2M}{r^{4}}+\frac{Q^2}{r^6})}~
 \bigg),  \label{BHmetricfactor} \\
F &=& -g\frac{2\sqrt 3 Q}{r^3} u_\mu dx^\mu \wedge dr,~~~A=(e A_{\mu}^{ext}-\frac{\sqrt 3 g Q}{r^2}u_{\mu})dx^{\mu}, \notag~~
\end{eqnarray}
with
\begin{equation}
\ell_c=\sqrt{\frac{1+U}{2}},~~U=\sqrt{1-8\alpha},~~u^v = \frac{1}{ \sqrt{1 - \beta_i^2} },~~u^i = \frac{\beta_i}{ \sqrt{1 - \beta_i^2} },~~P_{\mu \nu}= \eta_{\mu\nu} + u_\mu u_\nu, \label{velocity}
\end{equation}
where $\beta^i $, $M$, and $Q$ mean the velocity, black brane mass, and charge respectively
and $x^\mu=(v,x_{i})$ denote the boundary coordinates.
$P_{\mu\nu}$ corresponds to the projector onto spatial directions and
the boundary indices can be raised and lowered by the Minkowski metric
$\eta_{\mu\nu}$.
The outer horizon $r_{+}$ is located at the largest root satisfying $1-\frac{2M}{r^{4}}+\frac{Q^2}{r^6}=0$.
Notice that $\alpha$ should be smaller than $1/8$ from the definition of $U$. If not, the black brane factor in (\ref{BHmetricfactor}) becomes a complex number at the boundary. It is also worth noting that the AdS radius $\ell_c$ depends on the GB coupling $\alpha$.
The usual AdS radius, if we turn off the GB term, is given by $1$. For $\alpha>0$, the AdS radius $\ell_c$ becomes smaller than $1$, which implies that the t' Hooft coupling $\lambda$ of the dual gauge theory decreases according to the following relation $\ell_c^2 = \alpha' \sqrt{4 \pi \lambda}$. On the contrary, if $\alpha < 0$, the t' Hooft coupling increases. Therefore, it is interesting to study the dependence of various transport coefficients on the t' Hooft coupling.

Using the same method in Refs~\cite{Bhattacharyya:2008jc,Hur:2008tq,Erdmenger:2008rm,Banerjee:2008th,Tan:2009yg,
Torabian:2009qk,Hu:2010sn,Son:2009tf,Hu:2011ze}, we can define the following tensors
\begin{eqnarray}
\label{Tensors}
&&W_{IJ} = R_{IJ} + 4g_{IJ}+\frac{1}{6}\alpha L_{GB} g_{IJ}+\alpha H_{IJ}+\frac{1}{2g^2}\left(F_{IK}{F^{K}}_J +\frac{1}{6}g_{IJ}F^2\right),\\
&&W_A = \nabla_{B} {F^{B}}_{A}-\kappa_{cs} \epsilon_{ABCDE}F^{BC}F^{DE}~~.
\end{eqnarray}
When we take the parameters $\beta^i $, $M$, $Q$ and $A_{\mu}^{ext}$ as functions
of $x^\mu$ in~(\ref{rnboost}),
$W_{IJ}$ and $W_{A}$ become nonzero values proportional to the derivatives of
the parameters. From now on, we just focus on the case where these nonzero $W_{IJ}$ and $W_{A}$ are small and can be treated as the fluctuations around the background.
In the perturbative expansion, these nonzero terms can be considered as the source terms,
$S_{IJ}$ and $S_{A}$, of the next order equations.
In order to satisfy the gravitational and Maxwell equations even at the higher
orders
some extra gravitational and Maxwell field fluctuations must be added into (\ref{rnboost}).
In more details,
let us expand the parameters around $x^\mu=0$ up to the first order
\begin{eqnarray}
\beta_i&=&\partial_{\mu} \beta_{i}|_{x^\mu=0} x^{\mu},~~~M=M(0)+\partial_{\mu}
 M|_{x^\mu=0} x^{\mu},~~~Q=Q(0)+\partial_{\mu} Q|_{x^\mu=0} x^{\mu},\notag\\
A_{\mu}^{ext}&=&A_{\mu}^{ext}(0)+\partial_{\nu} A_{\mu}^{ext}|_{x^\mu=0} x^{\nu}, \label{Expand}
\end{eqnarray}
where we have assumed $\beta^i(0)=0$. Inserting the metric (\ref{rnboost}) and~(\ref{Expand})
into $W_{IJ }$ and $W_{A}$ does not make $W_{IJ }$ and $W_{A}$ zero, so
the rest of $-W_{IJ }$ and $-W_{A}$
should be considered as the source terms $S^{(1)}_{IJ }$ and $S^{(1)}_{A}$ of the first
order perturbation.
After fixing some gauge (the `background field' gauge in \cite{Bhattacharyya:2008jc})
\begin{equation}
G_{rr}=0,~~G_{r\mu}\propto u_{\mu},~~Tr((G^{(0)})^{-1}G^{(1)})=0,\label{gauge}
\end{equation}
and imposing the spatial $SO(3)$ symmetry which is also preserved in the background
metric~(\ref{rnboost}), the first order gravitational and Maxwell field fluctuations around $x^\mu=0$ can have the following form
\begin{eqnarray}\label{correction}
&&{ds^{(1)}}^2 = \frac{ k(r)}{r^2}dv^2 + 2 h(r)dv dr + 2 \frac{j_i(r)}{r^2}dv dx^i
 +\frac{r^2}{\ell_{c}^2} \left(\alpha_{ij} -\frac{2}{3} h(r)\delta_{ij}\right)dx^i dx^j, \\
&&A^{(1)} = a_v (r) dv + a_i (r)dx^i~~.\label{correction1}
\end{eqnarray}
Note that $a_r(r)$ in the gauge field fluctuations does not contribute to
field strength, thus the following gauge fixing $a_r(r)=0$ is natural.
As a result, solutions of the first order perturbation can be obtained
 from the vanishing $W_{IJ} = (\text{effect from correction})
  - S^{(1)}_{IJ }$ and $W_{A} = (\text{effect from correction}) - S^{(1)}_{A}$. Here, the `effect from correction' means the correction to $W_{IJ}$ and $W_{A}$ from (\ref{correction}) and (\ref{correction1}).

Since the Chern-Simons term does not affect the gravitational equations in (\ref{Tensors}), the first order gravitational equations are the same as the case without Chern-Siomns term. These first order gravitational equations $W_{IJ}=0$ are complicated, which have been already
written in the Appendix A in Ref~\cite{Hu:2011ze} explicitly.  We will not list these gravitational equations here again. However, the first order Maxwell equations become
\begin{eqnarray}\label{Maxwell}
&&W_v = \frac{f(r)}{r}\left\{ r^3 {a_v}' (r) + 4\sqrt 3 g Q h(r) \right\}'- S^{(1)}_{v}(r)=0~,\\\nn
&& W_r = - \frac{1}{r^3}\left\{  r^3 {a_v}'(r) +4\sqrt 3 g Q h(r)  \right\}'-S^{(1)}_r (r)=0~,\\\nn
&&W_i = \frac{1}{r} \left\{ r^3 f(r) {a_i}'(r) - \frac{2 \sqrt 3 g Q }{r^4} j_i (r) \right\}' -  S^{(1)}_i (r)=0~,
\end{eqnarray}
where
\begin{eqnarray}
&&S_v^{(1)}(r)=g\frac{2 \sqrt{3}}{r^3} \left(\partial _vQ+Q \partial _i\beta _i\right),
\nonumber\\&&S_r^{(1)}(r)=0,\nonumber\\&&S_x^{(1)}(r)=g \left(-\frac{\sqrt{3}}{r^3} \left
(\partial _xQ+Q \partial _v\beta _x\right)-\frac{1}{r} \frac{e}{g}
F^{\rm ext}_{vx} \right)-\kappa_{cs}\frac{16g\ell_{c}Q}{r^6}\left(\sqrt{3}
er^2F^{\rm ext}_{zy}-3gQ\partial _z\beta _y+3gQ\partial _y\beta _z \right)~.\nonumber\\\label{Sourceterm}
\end{eqnarray}
and $F^{\rm ext}_{vi}\equiv \partial_v A^{\rm ext}_i-\partial_i A^{\rm ext}_v$ is the external field strength tensor. In the above,
the prime means derivative with respect to $r$.  Compared with the case without Chern-Simons term, the Chern-Simons term affects the first order perturbative equations just through $S^{(1)}_i (r)$. Note that we write down only the $x$ component in (\ref{Sourceterm}) because
the other components, $y$ and $z$, can be easily obtained from the cyclic permutation of indices.

By solving all the first order perturbative equations, the undetermined functions in
(\ref{correction}) and (\ref{correction1}) are fixed to
\begin{eqnarray}
&&h(r) = 0,~~ k(r) = \frac{2}{3}r^3 \partial_i\beta^i,~~a_{v}(r)=0,\\\nn
 &&\alpha_{ij} =  \alpha(r)\left\{ (\partial_i \beta_j + \partial_j
 \beta_i )-\frac{2}{3} \delta_{ij}\partial_k \beta^k \right\},
\end{eqnarray}
where $\alpha(r)$ and its asymptotic expression are
\begin{eqnarray}
\alpha(r)&&= \int_{\infty }^{r}\frac{s^{3}-2\alpha s^{2}[s^{2}f(s)]^{^{\prime}}-
(r_{+}^3-2\alpha r_{+}^2(r^2f)'|_{r_{+}})}{-s+2\alpha [ s^{2}f(s)]^{^{\prime }}}\frac{1}{%
s^{4}f(s)}ds \notag\\
&&\approx \frac{\ell_c^2}{r}-\frac{1}{r^4}\frac{\alpha(r_{+}^6+12Q^2 \alpha
-16Mr_{+}^2 \alpha)}{r_{+}^3(1-\sqrt{1-8\alpha})\sqrt{1-8\alpha}}+O(\frac{1}{r})^5 ,
\end{eqnarray}
and we used the same frame and conditions as~\cite{Bhattacharyya:2008jc,Hu:2011ze}, (i.e., Landau frame, asymptotical AdS boundary condition and renormalization condition).
At the first order perturbation, the equations $W_i =0$ and $W_{ri} =0$
are explicitly written as
\begin{eqnarray}
&&\frac{r}{2} \left(\frac{j_i'(r)}{r^3}\right)'-\frac{\sqrt{3} Q}{g r^3}
a_i'(r)+\frac{8\alpha j_i(r)f'(r)}{r^3}+\frac{6\alpha j_i'(r)f(r)}{r^3}-
 \frac{2\alpha j_i'(r)f'(r)}{r^2}-\frac{2\alpha j_i''(r)f(r)}{r^2}=S_{r i}^{(1)}(r), \notag\\
&& \left(r^3 f(r) a_i'(r)-\frac{2 \sqrt{3} g Q}{r^4} j_i(r)\right)'=r S_i^{(1)}(r). \label{appeq1}
\end{eqnarray}
Since we concentrate only on the first order perturbation
we drop, for convenience, the superscript $(1)$
in $S_{r i}^{(1)}(r)$ and $S_i^{(1)}(r)$ from now on. After some algebra, the second order
differential equation of $j_i(r)$ reduces to~\cite{Hur:2008tq}
\begin{eqnarray}
j_i{}^{\prime\prime }(r)-(\frac{3}{r}-\frac{4\alpha
f'(r)}{-1+4\alpha f(r)}) j_i'(r)-(\frac{-12 Q^2+16r^7 \alpha
f(r)f'(r)} {r^8 f(r)(-1+4\alpha f(r))} )j_i(r)  = \zeta_i(r),
\label{appeqj}
\end{eqnarray}
where
\begin{equation}
\zeta_i(r)\equiv\big(-\frac{12 Q^2}{r^4 f(r)}
\frac{j_i\left(r_+\right)}{r_+^4}+2 r^2 S_{r i}(r)+\frac{2 \sqrt{3}
Q}{g r^4 f(r)} \int _{r_+}^rdx x S_i(x))/(1-4\alpha f(r)). \label{zetafu}
\end{equation}
This is the same as one without Chern-Simons term (see the appendix B in Ref~\cite{Hu:2011ze}) except the different $S_i(r)$.
Note that $\zeta_i(r)$ in (\ref{zetafu}) has the same divergence as one without Chern-Simons term, which provide the same contribution to $b_{2}(r)$ in $(B5)$ of \cite{Hu:2011ze} as $r$ goes to infinity.
Thus, $3x^2(2-1/\ell_{c}^2)\partial_{v}\beta_{i}$ should be added to cancel this divergence.
In other words, following the Ref~\cite{Hu:2011ze}, the exact form of $j_i(r)$
becomes
\begin{eqnarray}
j_i(r)&=&-r^4 f(r)\int _r^{\infty }dx x f(x)(1-4 \alpha f(x)) \zeta _i(x)
\int _x^{\infty }\frac{dy}{y^5 f(y)^2 (1-4 \alpha f(y))} \nonumber\\
&& +r^4 f(r) \left(\int _r^{\infty }\frac{dx}{x^5 f(x)^2 (1-4 \alpha f(x))}
\right) \Big(r^3(2-\frac{1}{\ell_{c}^2}) \partial _v\beta _i \nonumber\\
&&+\int _r^{\infty }dx \left[x f(x)(1-4 \alpha f(x)) \zeta _i(x)+3
x^2 (2-\frac{1}{\ell_{c}^2})\partial _v\beta _i\right]\Big).
\label{appjsol}
\end{eqnarray}
After inserting $S_i(r)$ in (\ref{Sourceterm}) into (\ref{appjsol}), the asymptotic behavior of $j_i(r)$ can be represented as
\begin{eqnarray}
j_x(r) &\approx& r^3 \partial _v\beta _x-\frac{8 \sqrt{3} Q \alpha}{5 r (-1+8 \alpha
+\sqrt{1-8 \alpha})} \frac{e}{g}F_{v x}^{\text{\rm ext}}+\frac{4 \alpha}
{r^2(-1+8 \alpha +\sqrt{1-8 \alpha})} \nonumber\\
&&\left(-Q^2 \frac{j_x\left(r_+\right)}{r_+^4}-\frac{Q}{2 r_+}
\left(\partial _xQ+Q \partial _v\beta _x\right)\right.
\left.+\frac{r_+ Q}{2 \sqrt{3}} \frac{e}{g}F_{v x}^{\text{\rm
ext}}\right)\nonumber\\
&&+\frac{4\sqrt{2} \alpha \sqrt{\sqrt{1-8\alpha}+1}\kappa_{cs}} {r^2(1-8 \alpha
-\sqrt{1-8 \alpha})} \left(\frac{Q^3g\sqrt{3}}{r^4_+}
\left(\partial _y\beta _z-\partial _z\beta _y \right)+\frac{
2Q^2e}{r^2_+} F_{z y}^{\text{\rm ext}})\right).\label{Asymji}
\end{eqnarray}
Near the outer horizon $r_+$, $j_i(r_+)$ is approximately
\begin{eqnarray}
\frac{j_x\left(r_+\right)}{r_+^4} &\approx&\frac{2 \left(2
r_+^6+Q^2\right) \partial _v\beta _x-Q \left(\partial _xQ+Q \partial
_v\beta _x\right)-\sqrt{3} r_+^2 Q
\frac{e}{g}F_{v x}^{\text{\rm ext}}}{8 M r_+^3}\nonumber\\
&+&\kappa_{cs}\frac{ 2\sqrt{3}Q^3g\ell_c (\partial
_z\beta _y-\partial_y\beta _z)-3 r_+^2 Q^2 e \ell_c(F_{z y}^{\text{\rm
ext}})}{2 M r_+^6}, \label{j at rp}
\end{eqnarray}
where the relationship between $\ell_{c}$ and $\alpha$ in (\ref{velocity}) has been used.
Here, the Chern-Simons effects proportional to $\kappa_{cs}$ are included in the last
terms in (\ref{Asymji}) and (\ref{j at rp}).
After integrating the second equation in (\ref{appeq1}) from
$r=r_+$ to $r$, we can get
\begin{eqnarray}
r^3 f(r) a_i'(r)-2 \sqrt{3} g Q
\left(\frac{j_i(r)}{r^4}-\frac{j_i\left(r_+\right)}{r_+^4}\right)=\int
_{r_+}^rdx x S_i(x). \label{appeq2i}
\end{eqnarray}
Integrating this one more time from $r$ to $\infty$, we finally obtain
\begin{eqnarray}
a_i(r) &=& \int_\infty^r dx \frac{1}{x^3f(x)}\left(2\sqrt{3}g
Q(\frac{j_{i}(x)}{x^4}-\frac{j_i(r_+)}{r_{+}^4})+ \int_{r_+}^x dy y
S_{i}(y)\right),
\end{eqnarray}
where the fact that $a_i(r)$ vanishes at infinity was used.
From this integral the asymptotic behavior of $a_i(r)$, especially $x$-component,
reduces to
\begin{eqnarray}
a_x(r) &\approx& \frac{\ell_{c}^2}{r} eF_{v x}^{\text{\rm
ext}}+\frac{\ell_{c}^2}{r^2} \left(\sqrt{3} gQ
\frac{j_x\left(r_+\right)}{r_+^4}+\frac{\sqrt{3}g}{2 r_+}
\left(\partial _xQ+Q \partial _v\beta _x\right)-\frac{r_+}{2} eF_{v
x}^{\text{\rm ext}}\right) \\
&&+\frac{\kappa_{cs}\ell_{c}^3}{r^2}\left(\frac{-6 g^2 Q^2(\partial _z\beta _y-\partial _y\beta _z)}{r_+^4 }+\frac{4 \sqrt{3} g e Q F_{z
y}^{\text{\rm ext}}}{r_+^2}\right). \nonumber
\end{eqnarray}
These asymptotic behaviors of $j_i(r)$ and $a_i(r)$ are
sufficient for evaluating the dual stress tensor and charge current at the boundary.

Now, we investigate the first order hydrodynamics of the dual conformal field theory
via AdS/CFT correspondence. Following the same procedure in Ref~\cite{Hu:2011ze},
the first order perturbative black brane solution determines the
first order stress tensor of the dual fluid $\tau _{\mu \nu}$~\cite{Myers:1987yn}
\begin{equation}
\tau_{\mu \nu}=\frac{1}{16\pi G}[\frac{2M}{\ell_{c}^3}(\eta_{\mu\nu} + 4 u_\mu u_\nu)-\frac{2r_{+}^2(r_{+}-8 \pi T \alpha)}{ \ell_{c}^3}\sigma_{\mu\nu}]=P(\eta_{\mu\nu} + 4 u_\mu u_\nu  ) - 2 \eta \sigma_{\mu\nu}, \label{StressTensor}
\end{equation}
where the counter term method was used~\cite{Balasubramanian:1999re,Emparan:1999pm,Mann:1999pc,Brihaye:2008xu,Cremonini:2009ih}.
The Hawking temperature is given by
\begin{equation}
T=\frac{(r^2f(r))'}{4 \pi}|_{r=r_{+}}=\frac{1}{2 \pi r_{+}^3}(4M-\frac{3Q^2}{r_{+}^2}), \label{Temperature}
\end{equation}
and the pressure and viscosity are read off
\begin{equation}
P=\frac{M}{8 \pi G \ell_{c}^3},~~~\eta=\frac{r_{+}^2(r_{+}-8 \pi T \alpha)}{16 \pi G \ell_{c}^3}, \label{ets}
\end{equation}
where $16 \pi G$ is temporally recovered for more convenient comparison.
As we expected, the Chern-Simons term does not affect the transport coefficients related to
the metric fluctuations.

Next, we consider the Chern-Simons effect on the current of the dual conformal field
theory. At the first order perturbation,
the charge current including the Chern-Simons effect is given by
\begin{equation}\label{current}
J^\mu = \lim_{r \rightarrow \infty} \frac{r^4}{\ell_{c}^4} \frac{1}{\sqrt{-\gamma}} \frac{\delta S_{cl}}{\delta \tilde A_\mu} =  \lim_{r \rightarrow \infty}  \frac{r^4}{\ell_{c}^4} \frac{N}{g^2} (F^{r \mu}+\frac{4\kappa_{cs} }{3}\epsilon ^{r\mu \rho \sigma \tau }A_{\rho
}F_{\sigma \tau })~~,
\end{equation}
where $\tilde A_\mu$ is the gauge field projected to the boundary.
After some complicate but direct algebras, the current near the zero point $x^{\mu}=0$
is given by
\begin{eqnarray}
J^\mu &=& J_{(0)}^\mu + J_{(1)}^\mu, \\ \label{first order current Component}
J_{(1)}^{v }&=&\kappa_{cs} \frac{ 8e^2}{3g^2}\left(F^{\rm ext}_{zy}A_{x}^{ext}
(0)+F^{\rm ext}_{xz}A_{y}^{ext}(0)+F^{\rm ext}_{yx}A_{z}^{ext}(0)
\right),\nonumber\\
J^{x}_{(1)} &=&\frac{1}{gl_{c}}\left\{-2\sqrt{3}Q\frac{j_{x}(r_{+})}{r_{+}^{4}}-%
\frac{\sqrt{3}\partial _{x}Q}{r_{+}}+\frac{er_{+}F^{\rm ext}_{vx}}{g}-\frac{\sqrt{3}Q}{%
r_{+}}\partial _{v}\beta _{x}\right\}\nonumber\\
&&+\kappa _{cs}\left(\frac{8e^2}{3g^{2}}(F^{\rm ext}_{yz}A_{v}^{ext}
(0)+F^{\rm ext}_{zv}A_{y}^{ext}(0)+F^{\rm ext}_{vy}A_{z}^{ext}(0))-\frac{8\sqrt{3}%
eF^{\rm ext}_{zy}Q}{gr_{+}^{2}}+\frac{12Q^{2}}{r_{+}^{4}}(\partial _{z}\beta
_{y}-\partial _{y}\beta _{z})\right)\nonumber\\
&=&\frac{1}{gl_{c}}\left\{-\frac{\sqrt{3}Q(2r_{+}^{6}+Q^{2})}{2Mr_{+}^{3}}\partial
_{v}\beta _{x}+(\frac{2\sqrt{3}Q^{2}}{8Mr_{+}^{3}}-\frac{\sqrt{3}}{r_{+}}%
)(\partial _{x}Q+Q\partial _{v}\beta _{x}) \right.\nonumber \\
&& \left.+(\frac{3eQ^{2}}{4gMr_{+}}+\frac{er_{+}}{g})F^{\rm ext}_{vx}\right\}+\kappa _{cs}\left((
\frac{-6Q^{4}}{r_{+}^{6}M}+\frac{12Q^{2}}{r_{+}^{4}})(\partial _{z}\beta
_{y}-\partial _{y}\beta _{z})\right.\nonumber \\
&& \left.+(\frac{3\sqrt{3}eQ^{3}}{gMr_{+}^{4}}-\frac{8\sqrt{3}eQ}{gr_{+}^{2}}%
)F^{\rm ext}_{zy}+\frac{8e^{2}}{3g^{2}}(F^{\rm ext}_{yz}A_{v}^{ext}(0)
+F^{\rm ext}_{zv}A_{y}^{ext}(0)+F^{\rm ext}_{vy}A_{z}^{ext}(0))\right)\nonumber\\
&=& \frac{1}{g \ell_{c}} \left\{-2 \sqrt{3} Q \frac{j_{\beta }
\left(r_+\right)}{r_+^4}\partial_v \beta_x +\left(
-2 \sqrt{3} Q \frac{j_Q\left(r_+\right)}{r_+^4}-\frac{\sqrt{3}}{r_+}\right)
 \left(\partial_x Q+Q\partial_v\beta_x \right) \right.\nonumber\\
&& \left.+\left(-2 \sqrt{3} Q \frac{j_F\left(r_+\right)}{r_+^4}+\frac{e}{g}r_
+\right) F ^{\rm ext}_{vx}\right\}+\kappa_{cs} \left(
-\frac{\sqrt{3}eQF^{\rm ext}_{zy}}{gMr_{+}^4}(Q^2+4r_{+}^6)+\frac{6Q^2}{M}
(\partial_z \beta_y-\partial_y \beta_z)\right.\nonumber\\
&&\left.+\frac{8e^{2}}{3g^{2}}(F^{\rm ext}_{yz}A_{v}^{ext}(0)
+F^{\rm ext}_{zv}A_{y}^{ext}(0)+F^{\rm ext}_{vy}A_{z}^{ext}(0)) \right),
\end{eqnarray}
where (\ref{j at rp}) has been used and the zeroth order boundary current is
\begin{equation}
J_{(0)}^{\mu} = \frac{2\sqrt 3 Q}{g \ell_{c}^3} u^\mu :=nu^\mu .
\end{equation}
$j_\beta(r_+)$, $j_Q(r_+)$, and $j_F(r_+)$ are values of each function at the horizon
\begin{eqnarray}
\frac{j_{\beta }\left(r_+\right)}{r_+^4}=\frac{2 \left(2
r_+^6+Q^2\right)}{8 M
r_+^3}~~,~~\frac{j_Q\left(r_+\right)}{r_+^4}=-\frac{Q}{8 M r_+^3}~~,
~~\frac{j_{F}\left(r_+\right)}{r_+^4} =-\frac{e}{g}\frac{\sqrt{3}
Q}{8 M r_+}~~ \label{jis}  .
\end{eqnarray}
Here we explicitly present $j_{F,Q,\beta}(r_+)$ just for the convenient comparison with results without Chern-Simons term.

Following the procedure discussed in Ref~\cite{Bhattacharyya:2008jc}, the global current defined on the whole boundary can be obtained by rewriting (\ref{first order current Component}) as a covariant form like $\partial_{v}\beta_{i} \to u^{v}\partial_{v} u_{\mu}$. As a result, the first order global charge current can be represented as
\begin{eqnarray}
J_{(1)}^{\mu }=-\kappa P^{\mu \nu }\partial _{\nu}(\frac{\mu }{T}%
)+\sigma _{E}E^{\mu }+\sigma _{B}B^{\mu }+\xi \omega ^{\mu }+\ell \epsilon ^{\mu
\nu \rho \sigma }F_{\rho \sigma }^{\text ext}A_{\nu}^{\text ext}, \label{first order current}
\end{eqnarray}
where
\begin{eqnarray} \label{firstcoeff}
\kappa &=&  \frac{\pi ^2 T^3 r_+^7}{4 g^2M^2 \ell_{c}},~\sigma_{E} =\frac{\pi ^2 e^2 T^2 r_+^7}{4g^2 M^2 \ell_{c}},
~\sigma _{B}=-\frac{\sqrt{3} \kappa_{cs} e Q (3r_{+}^4+2M)}{gMr_{+}^2},~\xi=\frac{6 \kappa_{cs} Q^2}{M},\nonumber\\
\ell&=&\frac{4 \kappa_{cs} e^2}{3 g^2},~E^{\mu}=u^{\lambda } F ^{\rm ext}{}_{\lambda }{}^{\mu },
~B^{\mu }=\frac{1}{2}\epsilon ^{\mu \nu \rho \sigma}u_{\nu}F_{\rho \sigma}^{ext},~\omega ^{\mu }=\epsilon ^{\mu \nu \rho \sigma }u_{\nu}\partial _{\rho}u_{\sigma},
\end{eqnarray}
and the chemical potential $\mu$ is defined as~\cite{Hur:2008tq,Erdmenger:2008rm}
\begin{eqnarray}\label{chemical potential}
\mu = A_{v} (r_+) - A_{v} (\infty) ~~.
\end{eqnarray}
Following the same discussions in Refs \cite{Hur:2008tq,Hu:2011ze}, we can find that its first order expression becomes
 \begin{eqnarray}
 \mu = 
 \frac{\sqrt 3 gQ(x)}{  r_+ ^2 (x)}~~,
 \end{eqnarray}
where $Q$ and $r_+$ are functions of coordinates.
After setting $16 \pi G =1$ and taking appropriate values for $g$ and $e$,
we can obtain the transport coefficients $\sigma _{B}$ and $\xi$ consistent with the results in~\cite{Erdmenger:2008rm,Banerjee:2008th,Son:2009tf}. In this case, the definition of chemical potential~(\ref{chemical potential}) is coincident with the thermodynamic relation $\mu=(\partial \epsilon /\partial n)_{s}$ because
the first law of thermodynamics $d\epsilon=Tds+\mu dn$ is satisfied with $\epsilon=\frac{3M}{8\pi Gl_{c}^{3}}$,~$s=\frac{r_{+}^{3}}{4Gl_{c}^{3}}$, $T=\frac{1}{2 \pi r_{+}^3}(4M-\frac{3Q^2}{r_{+}^2})$, $\mu=\frac{\sqrt 3 gQ}{  r_+ ^2}$, and $n=\frac{2\sqrt 3 Q}{g \ell_{c}^3}$.

In the charge current (\ref{first order current}), the first interesting thing is to find out the
dependence of the Gauss-Bonnet coupling $\alpha$ for the comparison with the Einstein case $
\alpha=0$. We can easily see that the charge diffusion constant $\kappa$ and electric
conductivity $\sigma_{E}$ depend only on $\alpha$, while the other three coefficients $\sigma_
{B}$, $\xi$ and $l$ are related to Chern-Simons term.
Remember that the GB coefficient $\alpha$ should be smaller than $1/8$. Especially for small
$\alpha$, the diffusion constant and the electric conductivity can be represented as the
perturbative form
\begin{eqnarray}
\kappa &=&  \frac{\pi ^2 T^3 r_+^7}{4 g^2M^2} \left( \frac{}{} 1 + \alpha + {\cal O} (\alpha^2) \right), \nonumber \\
\sigma_{E} &=&\frac{\pi ^2 e^2 T^2 r_+^7}{4g^2 M^2} \left(  \frac{}{} 1 + \alpha + {\cal O} (\alpha^2) \right) .
\end{eqnarray}
From these, we can see that the diffusion constant and the electric conductivity decrease
with the t' Hooft coupling because it decreases as  $\alpha$ increases.

In the charge current (\ref{first order current}), there are three transport coefficients related
to the Chern-Simons term. The coefficients of two terms, $\sigma_{B}$ and $\xi$, represent
the chiral magnetic conductivity and chiral vortical effect respectively, which were widely
investigated
by many authors~\cite{Erdmenger:2008rm,Banerjee:2008th,Son:2009tf,Hu:2011ze,Kalaydzhyan:2011vx,Kharzeev:2004ey,Fukushima:2008xe,Rubakov:2010qi,Landsteiner:2011iq,Amado:2011zx}.
The remaining coefficient $l$ of the last term
in (\ref{first order current})  was discussed in the
STU black brane background~\cite{Kalaydzhyan:2011vx}. Following \cite{Kalaydzhyan:2011vx}, this
term does not contribute the first order hydrodynamic transport coefficients but generate the second order one.
In this paper, we will
show that the last term can provide an additional effect to the first order hydrodynamic
coefficient. From now on, we call this term a new term for emphasizing the new effect on the
first order hydrodynamic coefficient, especially the chiral magnetic conductivity.


The new term is gauge-dependent like the gauge-dependent topological gluon current
\cite{Kharzeev:2004ey,Fukushima:2008xe}. However, its divergence still preserves
the $U(1)$ gauge invariance and is proportional to $E^{\mu}_{ext} B_{\mu}^{ext}$.
This new term as well as other two anomalous terms related to $\omega^{\mu}$ and $B^{\mu}$,
which are related to the nontrivial topological gluon configuration for QCD~\cite{Kharzeev:2004ey}, break the charge current conservation.
It is worth noticing that in the two flavor case the addition of the so-called Bardeen counter
term makes the vector current conserved but the axial current is not still conserved \cite
{Kalaydzhyan:2011vx}.
To understand an additional effect of the new term,
we need to generalize our one flavor model to the two flavor case by adding an additional U(1) gauge field \cite{Lee:2009bya}.
In this case, only the
change of the background geometry is the replacement of the black brane charge in (\ref
{BHmetricfactor}), $Q^2 \to 2 Q^2 $ where $2$ means the number of flavor \cite{Lee:2009bya}.
After adding the Bardeen term \cite
{Kalaydzhyan:2011vx,Rubakov:2010qi} or extra Chern-Simons current \cite{Landsteiner:2011iq} to the generalized two flavor model, we can easily prove the vector current conservation at the leading order.
Furthermore, if we also take into account the higher gradient corrections of the Bardeen term
which were not considered in \cite{Kalaydzhyan:2011vx},
the vector current conservation law is preserved even at the higher order perturbations regardless of the boundary value of the axial vector field. In spite of the Bardeen term
the axial current conservation still remain broken. In that sense,
the new term we considered provides an additional effect only to the axial current, which
implies
that the chiral magnetic conductivity of the axial current can be different from that of
the vector current by this additional effect.

To understand that the new term can provide an additional contribution to the axial
magnetic conductivity at the first order perturbation,
we take $A_{v}^{\text ext}$ to be a non-zero value at the origin ($x_{\mu} = 0$).
In the previous works~\cite{Erdmenger:2008rm,Banerjee:2008th,Son:2009tf,Hu:2011ze,Kalaydzhyan:2011vx}, $A_{\mu}^{\text ext}$ is zero at the origin and
slowly deviated from it near the origin.
In the holographic model \cite{Amado:2011zx,Lee:2009bya}, the non-zero $A_{v}^{\text ext}$ corresponds to the chemical potential of the dual field theory, which plays an important role to investigate the thermodynamics of the dual gauge theory.
In the case of the non-zero $A_{\mu}^{\text ext}$, the axial current ${J}_{(1)}^{\mu }$
can be simplified to
\begin{eqnarray}
{J}_{(1)}^{\mu}
&=&-\kappa  P^{\mu  \nu } \partial _{\nu }\frac{\mu }{T}+\sigma_{E} u^{\lambda } F ^{\rm ext}{}_{\lambda }{}^{\mu }+\sigma _{B}B^{\mu }+\xi \omega ^{\mu }, \label{first order current1}
\end{eqnarray}
which is the same form as the vector current~\cite{Hu:2011ze}.
Only the difference of the axial current from the vector one comes from the chiral magnetic
conductivity. If we set $A_{\mu}^{\text ext}=\{C,0,0,0\}$, the chiral magnetic conductivity of
the axial current is given by
\begin{eqnarray}		\label{cmconductivity}
\sigma _{B}=\frac{\sqrt{3} \kappa_{cs} e Q (3r_{+}^4+2M)}{gMr_{+}^2}+ 2  \ell C ,
\end{eqnarray}
where the last term is the additional contribution from the new term.
This result is consistent with that in~\cite{Amado:2011zx} where the anomalous magnetic
conductivity was discussed by using the Kubo formula.

\section{Conclusion and discussion}

In this paper, we considered the first order pertubative black brane solution in the  Maxwell-Gauss-Bonnet gravity and extracted, following the Gauge/Fluid correspondence, the hydrodynamic informations of the dual conformal field theory such as the stress tensor and charge current. Especially, we systematically investigated the Gauss-Bonnet and the Chern-Simons effects on the first order hydrodynamics.

Interestingly, the asymptotic geometry of the Gauss-Bonnet gravity still remains
as the AdS space with a different AdS radius, which implies that the dual gauge theory
is another conformal theory with a different 't Hooft coupling.
Although the stress tensor has the same form as that without the Chern-Simons
term~\cite{Hu:2011ze}, we showed that the hydrodynamic coefficients corresponding to the diffusion constant
and electric conductivity crucially depend on the Gauss-Bonnet coupling $\alpha$.
In the dual gauge theory point of view, the $\alpha$ dependence of these hydrodynamic
coefficients showed that they increase with $\alpha$. In other words,
since the 't Hooft coupling decreases with $\alpha$,  the charge diffusion constant
and electric conductivity decrease as the 't Hooft coupling increases.
In addition, we also studied the Chern-Simons effect on the hydrodynamic transport coefficients. We found that if the time-component of the external field $A^{ext}_{\mu}$ does not vanish,
the new term related to the axial charge current can provides an additional effect to the chiral magnetic conductivity. This is consistent with the result obtained by the different way, so called the Kubo's formula \cite{Amado:2011zx}. This result indicate that the chiral magnetic conductivity
of the axial current can be different from that of the vector current. So, it would be very interesting
to check this difference through the experiments.

Related to the new anomalous term of the axial current, it is also interesting to take into account a general $A_{\mu }^{ext} (x)$
because it may also provide the extra unexpected effects to the hydrodynamic transport
coefficients. However, it should be noticed that we are careful when choosing $A_{\mu }^{ext} (x)$ because there is a restriction from the Landau frame $u_{\mu}J_{(1)}^{\mu}=0$.
The effects and underlying physics related to this new term with the general ansatz are still unclear. Therefore, it seems to be important to investigate their effects in depth
like the other two anomalous terms
$\omega^{\mu}$ and $B^{\mu}$. We will leave these issues as the future works.

\section{Acknowledgements}

Y.P Hu thanks Prof. Rong-Gen Cai, Dr. Zhang-Yu Nie, Peng Sun, Jian-Hui Zhang for useful discussions. He also thanks a lot for Dr. F. Pena-Benitez's useful comments. Particularly, he thanks a lot for the CQUeST's hospitality during his visiting CQUeST, Sogang University, Korea. This work is supported by China Postdoctoral Science Foundation under Grant No.20110490227 and National Natural Science Foundation of China (NSFC) under grant No.11105004, and also
supported partially by grants from NSFC, No. 10975168 and No. 11035008.
C. Park was supported by the National Research Foundation of Korea(NRF) grant funded by
the Korea government(MEST) through the Center for Quantum Spacetime(CQUeST) of Sogang
University with grant number 2005-0049409 and also by Basic Science Research Program through the National Research Foundation of Korea(NRF) funded by the Ministry of
Education, Science and Technology(2010-0022369).



\end{document}